\documentclass{article}

\usepackage{booktabs,longtable,pdflscape,array,ragged2e}
\usepackage[utf8]{inputenc}
\usepackage[T1]{fontenc}
\usepackage[colorlinks=true,allcolors=blue,hypertexnames=false]{hyperref}
\usepackage{url}
\usepackage{booktabs}
\usepackage{microtype}
\usepackage{xcolor}
\usepackage{graphicx}
\usepackage{amsmath,amssymb,amsthm,amsfonts,bm}
\usepackage{algorithm}
\usepackage{algorithmic}

\usepackage{natbib}

\newcommand{\cA}{\mathcal{A}}
\newcommand{\bbeta}{\boldsymbol{\beta}}
\newcommand{\bgamma}{\boldsymbol{\gamma}}
\newcommand{\bx}{\boldsymbol{x}}
\newcommand{\bX}{\boldsymbol{X}}
\newcommand{\bz}{\boldsymbol{z}}
\newcommand{\bZ}{\boldsymbol{Z}}
\newcommand{\bzero}{\boldsymbol{0}}
\newcommand{\bone}{\boldsymbol{1}}

\newcommand{\bSigma}{\boldsymbol{\Sigma}}
\newcommand{\ind}{\mathbb{I}}

\theoremstyle{definition}

\newtheorem{definition}{Definition}

\title{Learning Interpretable Point-Based Clinical Risk Scores via Direct Optimization}

\author{\scriptsize Ying Cui$^1$, Albert M Li$^2$, Vivek Charu$^3$, Yeon-Mi Hwang$^4$, Tina Hernandez-Boussard$^4$, Lu Tian$^{1}$\\
\scriptsize $^1$ Department of Biomedical Data Science, Stanford University;\\
\scriptsize $^2$ Decatur High School;\\
\scriptsize $^3$ Department of Pathology, Stanford University School of Medicine;\\
\scriptsize $^4$ Division of Computational Medicine, Department of Medicine, Stanford University.\\
\scriptsize Correspondence email: lutian@stanford.edu.
}
\date{}

\begin{document}

\maketitle

\begin{abstract}
Many clinical risk scores are deployed as additive rules with nonnegative integer points assigned to relevant binary predictive features. These integer weights not only make the score easier to use in practice but also promote sparsity in the resulting prediction model. Such risk scores are often derived by first fitting a regression model and then rounding the estimated coefficients to the nearest integer after appropriate scaling. This approach is computationally fast but does not guarantee optimality of the resulting score. Alternatively, one may search over all possible integer weights to directly optimize a value function by posing the problem as an integer programming task. However, the associated computational burden can be substantial, especially when the value function is nonconcave or even discontinuous. In this paper, we develop new machine learning algorithms that employ a flexible greedy optimization strategy to learn such additive scoring directly under explicit and sensible optimality objectives. We apply the proposed method to a large electronic health record (EHR) cohort in Epic Cosmos to construct an integer-weighted comorbidity score for measuring the risk of post-discharge mortality. We also conduct a simulation study to examine the finite-sample operating characteristics.
\end{abstract}



\section{Introduction}

Interpretable and easy to use point-based scoring systems remain essential tools for clinical decision making. A classical example is the Charlson Comorbidity Index (CCI), which converts a collection of comorbidity indicators into an additive integer score for prognosis \citep{charlson1987}. The appeal of this format is practical rather than purely statistical. Specifically, each condition contributes a visible number of points, larger totals correspond to greater disease burden, and the scoring rule can be computed without storing a full model. Other examples include the CHADS$_2$ score for stroke risk in atrial fibrillation \citep{gage2001}, the APACHE II score for ICU mortality risk \citep{knaus1985}, and the SOFA score for dysfunction across six organ systems \citep{vincent1996}. 

To construct such a score, researchers often begin with a set of candidate conditions and a desire to preserve monotonicity, sparsity, and a small set of allowable integer weights. Standard regression does not solve this problem directly \citep{ustun2016}. A logistic or survival model with continuous coefficients is usually fit first, and its estimated coefficients are then rounded or grouped into a small number of integers. This post hoc discretization is convenient, but it does not guarantee good performance. 
These limitations have motivated methods that learn interpretable scoring systems directly under explicit structural and deployability constraints. {For example, \citet{ustun2019learning} developed RiskSLIM for sparse small-integer risk scores using mixed-integer optimization of logistic loss, which optimizes a surrogate loss rather than the eventual performance of the discretized score itself. More recently, \citet{ruhrbergestevez2026agentscore} studied unit-weighted clinical checklists built from LLM-generated binary rules, focusing on checklist construction with simple 0/1 weights rather than learning more general integer valued scores.}

{To address these limitations, we propose a method for directly learning point-based risk scores with a discrimination-based objective over any finite discrete coefficient space.} A motivating example is the construction of a comorbidity risk score as a linear combination of binary comorbidity indicators for predicting short- and medium-term mortality after inpatient admission. In that setting, the score should satisfy three goals. First, coefficients should be nonnegative so that the addition of a comorbidity cannot decrease the risk score. Second, the allowed coefficient values should be restricted to a small set of integers so that the resulting score is easy to use and interpret clinically. Third, the optimization target should reflect discriminatory performance, rather than rely on a continuous approximation that is later rounded.

The main contributions are as follows.
\begin{itemize}
    \item 
    We formulate direct learning of interpretable point-based risk scores as optimization over discrete, nonnegative coefficient spaces.
    \item We develop a basic greedy improvement algorithm that learns the score by directly maximizing an empirical pairwise concordance objective for binary outcomes.
    \item We extend this algorithm with look-ahead variants based on reinforcement learning, yielding efficient search strategies guided by long-term gains.
    \item We define a benchmark based on logistic regression with tuned coefficient rounding and evaluate all methods in simulations and an EHR comorbidity scoring application.
    \item We study model complexity through null AUC optimism and effective search complexity, suggesting how it changes with predictor dimension and coefficient levels.
\end{itemize}

\section{Motivating EHR Comorbidity Scoring Application}
\label{sec:motivation}

\subsection{Objective}

The CCI and related instruments illustrate the basic design of a comorbidity score, in which integer points are assigned to selected conditions and then summed into a burden index \citep{charlson1987}. Such scores are especially attractive in EHR research because a low-dimensional summary of baseline disease burden can be reused across outcomes, subgroups, and institutions. The core monotonicity requirement is intuitive, since adding a documented condition should not reduce the estimated risk score.

Let $X_j \in \{0,1\}$ indicate whether comorbidity $j$ is present. The target score is
\begin{equation}
S = \sum_{j=1}^{p} \beta_j X_j,
\label{eq:score}
\end{equation}
where the coefficient vector $\bbeta = (\beta_1,\ldots,\beta_p)^T$ is restricted to a discrete nonnegative action set $\cA$. For a CCI-like score, a natural choice is $\cA = \{0,1,\cdots, L\}$ for an appropriate $L.$ The binary set $\{0,1\},$ i.e., $L=1,$ is a useful special case. 

\subsection{Motivating Example}

The real data analysis uses the 1\% SneakPeek sample from Epic Cosmos, a large EHR research network assembled from participating health systems across the United States \citep{noel2023}. The target population consists of adult patients with an inpatient admission between March~31,~2015 and March~31,~2020. To improve ascertainment of baseline disease burden, we retain patients with at least two healthcare encounters before the index admission spanning at least 365 days. We exclude zero-day stays, pregnancy-related admissions, and repeat qualifying admissions, keeping only the first eligible inpatient encounter per patient.

We focus on $40{,}622$ patients older than 65 years and study 180-day mortality as the primary endpoint. The analysis is based on a training set of $24{,}373$ patients ($60\%$) and a test set of $16{,}249$ patients ($40\%$). Candidate predictors include 22 disease-condition indicators derived from the AHRQ CCSR ICD-10-CM diagnosis categories, as listed in the Appendix D, together with age and sex. Age is represented by the three indicators $\ind\{\mathrm{Age}\in[65,75)\}$, $\ind\{\mathrm{Age}\in[75,85)\}$, and $\ind\{\mathrm{Age}\in[85,100]\}$, and sex is represented by $\ind\{\mathrm{Sex}=\mathrm{Male}\}$, where $\ind(\cdot)$ is a binary indicator function. Together, these binary indicators form the components of \eqref{eq:score} in the real data application.

This analysis is intended to evaluate whether the proposed discrete optimization procedures can produce a clinically interpretable score while retaining predictive discrimination close to that of a continuous logistic benchmark. Accordingly, all methods are trained on the same set of binary predictors and compared on the held-out test set using 180-day mortality discrimination.

\section{Method}
\label{sec:method}

\subsection{Direct optimization over discrete coefficient spaces}

Suppose we observe $n$ independent pairs $(Y_i,\bX_i)$, where $Y_i$ is a binary outcome and $\bX_i = (X_{i1},\ldots,X_{ip})^\top$ is a vector of binary predictors. For a candidate coefficient vector $\bbeta$, subject $i$ receives score
\[
S_i(\bbeta) = \bbeta^\top \bX_i.
\]
The coefficients are constrained to lie in a finite ordered set $\cA=\{0, 1, \cdots, L\}$. This discrete set enforces interpretability and caps the maximum point contribution of a single comorbidity.

Let $n_1 = \sum_{i=1}^n \ind(Y_i=1)$ and $n_0 = \sum_{i=1}^n \ind(Y_i=0)$ denote the number of cases and controls, respectively. We use the objective
\begin{equation}
\ell(\bbeta)
=
\frac{1}{n_1n_0}
\sum_{i:Y_i=1}\sum_{j:Y_j=0}
\left[\ind\!\left\{S_i(\bbeta) > S_j(\bbeta)\right\}+\frac{1}{2}\ind\left\{S_i(\bbeta)=S_j(\bbeta)\right\}\right],
\label{eq:target}
\end{equation}
which measures empirical case--control pairwise concordance. This criterion is equivalent to the area under the ROC curve and directly rewards separation between case and control under the final discrete score \citep{hanley1982,pepe2000,pepe2006}. Our objective is to find a coefficient vector $\bbeta$ maximizing the objective function \eqref{eq:target}. 

The nonnegativity restriction has an important clinical interpretation. If $\beta_j \ge 0$ for every comorbidity, then documenting an additional condition cannot decrease the score. This matches the intended semantics of comorbidity burden and prevents pathological sign reversals that may arise when highly correlated diagnoses compete in an unconstrained model. The constraint on the coefficient vector also introduces regularization from the optimization perspective. It is important to understand the strength of this regularization. In Appendix A, we quantify this regularization through standardized optimum bias, which is approximately proportional to $p$ and $\log(L+1)$ and can help guide the choice of $p$ and $L$ relative to the sample size of the training data.

\subsection{Greedy target-improvement algorithm}

Our current primary method is the conventional greedy search described in Algorithm \ref{alg:greedy}. Starting from $\bbeta^{(0)} = \bzero$, the algorithm evaluates every one-coordinate update obtainable by setting coefficient $j$ to action value $a \in \cA$. For each candidate state
\[
\bbeta^{(t,j,a)} = (\beta_1^{(t)},\ldots,\beta_{j-1}^{(t)}, a, \beta_{j+1}^{(t)},\ldots,\beta_p^{(t)}),
\]
we compute the immediate improvement
\begin{equation}
r_0(j,a) = \ell\!\left(\bbeta^{(t,j,a)}\right) - \ell\!\left(\bbeta^{(t)}\right).
\label{eq:gain}
\end{equation}
The next action is chosen as the candidate with the largest positive gain.  When multiple updates achieve the same improvement, the algorithm uses a deterministic tie-breaking rule. The current priority order favors actions that modify already-active coefficients, then predictors whose prevalence is closer to $0.5$, and finally smaller movements from the current coefficient value. This rule stabilizes the path and helps avoid arbitrary differences across equivalent updates.

\begin{algorithm}[t]
\caption{Greedy target improvement for discrete comorbidity scoring}
\label{alg:greedy}
\footnotesize
\begin{algorithmic}[1]
\REQUIRE Action set $\cA$, objective $\ell(\bbeta)$, initial coefficients $\bbeta^{(0)}=\bzero$
\STATE $t \gets 0$
\WHILE{true}
    \FOR{$j=1$ to $p$}
        \FOR{each $a \in \cA$}
            \STATE Set $\bbeta^{(t,j,a)} \gets \bbeta^{(t)}$ and replace its $j$th entry by $a$
            \STATE Compute $r_0(j,a) \gets \ell(\bbeta^{(t,j,a)}) - \ell(\bbeta^{(t)})$
        \ENDFOR
    \ENDFOR
    \STATE Choose $(j^\star,a^\star) \in \arg\max_{j,\,a\in\cA} r_0(j,a)$ using deterministic tie-breaking
    \IF{$r_0(j^\star,a^\star) \le 0$}
        \STATE \textbf{break}
    \ENDIF
    \STATE Update $\bbeta^{(t+1)} \gets \bbeta^{(t,j^\star,a^\star)}$
    \STATE $t \gets t+1$
\ENDWHILE
\STATE \textbf{return} $\hat{\bbeta} \gets \bbeta^{(t)}$
\end{algorithmic}
\normalsize
\end{algorithm}

The algorithm is easy to implement efficiently. If the current score vector $(S_1(\bbeta^{(t)}),\ldots,S_n(\bbeta^{(t)}))$ is cached, then each candidate score can be updated by
\[
S_i\!\left(\bbeta^{(t,j,a)}\right)
=
S_i\!\left(\bbeta^{(t)}\right) + \left(a-\beta_j^{(t)}\right)X_{ij}.
\]
The main computational burden is therefore evaluating pairwise concordance across candidate updates (see Remark 1), not manipulating the coefficient vector itself.

\paragraph{Model tuning}
The greedy algorithm \ref{alg:greedy} terminates when no further update improves $\ell(\bbeta)$. Because all coefficients are restricted to set $\cA$, overfitting to the training data is expected to be less severe than in a standard regression, and the final weights $\hat{\bbeta}$ can be used directly to construct the risk score. Nevertheless, one can also use cross-validation to determine an earlier stopping point and thereby further guard against overfitting. In some applications, the action set $\cA$ can also be selected adaptively by cross-validation.

\paragraph{Remark 1}
Because the score $S_i(\bbeta)$ is discrete, the objective in \eqref{eq:target} can be evaluated much faster than by enumerating all $N_1N_0$ case--control pairs. Suppose the score takes only $M$ ordered values, denoted by $s_1 < \cdots < s_M$. Let $n_{1m}$ and $n_{0m}$ be the numbers of cases and controls, respectively, with score being $s_m$. Then pairwise concordance can be computed from these tabulated frequencies: 
\[
\ell(\bbeta)
=
\frac{1}{n_1n_0}
\sum_{m=1}^M n_{1m}\left(\sum_{\ell < m} n_{0\ell} + \frac{1}{2}n_{0m}\right).
\]
This objective can be obtained in $O(Mn)$ time, which is especially attractive when $M \ll n$.

There are two major extensions of the base algorithm. The first extension of the base algorithm is a \emph{local-step greedy} procedure that only considers adjacent moves in an ordered action set $\cA = \{0, 1, \cdots, L\}$. If the current coefficient satisfies $\beta_j^{(t)} = a$, then only members in $\{a-1, a+1\}\cap\cA$ are considered as candidate updates. This restriction reduces the search burden and may produce smoother coefficient paths.

The second extension of the base algorithim is a \emph{look-ahead} method. Although the greedy method is computationally attractive, it can be myopic: selecting the update that maximizes the immediate gain in \eqref{eq:gain} may miss a move that yields a better final solution after subsequent coefficient updates. Following the core idea of reinforcement learning \citep{sutton2018}, a natural refinement is to evaluate a candidate update by its ``look-ahead'' value rather than by its one-step improvement alone. Specifically, after forcing coordinate $j$ to take value $a$ at iteration $t$, let ${\bbeta}_{r_0}^{(t,j,a)}$ denote the final coefficient vector obtained by rerunning Algorithm~\ref{alg:greedy} from the initial state $\bbeta^{(0)}=\bbeta^{(t,j,a)}$:
\[\bbeta^{(t,j,a)} \xrightarrow{\text{Algorithm \ref{alg:greedy}}} \bbeta_{r_0}^{(t,j,a)} .\]
 We then define the gain of the candidate move by
\begin{equation}
r_1(j,a) = \ell\!\left(\bbeta_{r_0}^{(t,j,a)}\right)-\ell(\bbeta^{(t)})
\label{eq:lookahead-gain}
\end{equation}
This criterion asks not whether the move is best immediately, but whether it leads to the best terminal score after the remaining greedy steps are taken. In summary, $r_1(j, a)$ is expected to be a sensible estimator for the value function of candidate update $(j, a)$ in the reinforcement learning.  The update with the largest $r_1(j, a)$ is then chosen.

The main drawback of the \emph{look-ahead} algorithm is computational cost. The look-ahead method evaluates each candidate update by using the resulting coefficient vector as the initial state for the conventional greedy procedure. Evaluating $r_1(j,a)$ exactly therefore requires rerunning Algorithm~\ref{alg:greedy} for every candidate update $(j,a)$ under consideration, which can be expensive when $p$ or $L$ is large. In practice, however, several approximations can substantially reduce the burden. One option is to compute $r_1(j,a)$ only for the top $K$ candidate moves ranked by the one-step gain $r_0(j,a)$ in \eqref{eq:gain}, which is relatively cheap to evaluate. Another option is to use a truncated look-ahead in which Algorithm~\ref{alg:greedy} is run for only a fixed number of additional iterations when approximating $r_1(j,a)$. Alternatively, one can cache the greedy continuation from previously visited intermediate coefficient vectors. When the same intermediate state reappears, its downstream value can be retrieved rather than recomputed. These heuristics preserve the basic idea of accounting for downstream improvement while keeping the procedure computationally feasible. Similarly, we can develop a \emph{local-step look-ahead} algorithm following the same rationale as for the \emph{local-step greedy} algorithm.

\paragraph{Remark 2.} In principle, this look-ahead rule can be extended further by defining a better approximation to the ``value'' of a candidate update in reinforment learning:
\begin{equation}
r_k(j,a) = \ell\left(\bbeta_{r_{k-1}}^{(t,j,a)}\right) - \ell\left(\bbeta^{(t)}\right),  k=2, 3, \cdots, 
\label{eq:general-lookahead-gain}
\end{equation}
where $\bbeta_{r_{k-1}}^{(t,j,a)}$ denotes the final coefficient vector obtained by rerunning Algorithm~\ref{alg:greedy} from the initial state $\bbeta^{(0)} = \bbeta^{(t,j,a)}$, but with the one-step gain $r_0(j,a)$ in Step~6 replaced by the look-ahead gain $r_{k-1}(j,a)$ in (\ref{eq:lookahead-gain}) for $k=2$ and (\ref{eq:general-lookahead-gain}) for $k>2$. The main cost of this recursive refinement is its computational budern.

\subsection{Reference methods}

The reference comparator mirrors the classical score-construction workflow. We first fit logistic regression to the binary endpoint and obtain continuous coefficients $\beta_1^{\mathrm{logit}},\ldots,\beta_p^{\mathrm{logit}}$. For a scaling factor $\Delta > 0$, we define rounded integer weights
\[
w_k(\Delta) = \operatorname{round}\!\left(\beta_k^{\mathrm{logit}} / \Delta\right), \qquad k=1,\ldots,p,
\]
and corresponding score $S_i(\Delta) = \sum_{k=1}^{p} w_k(\Delta) X_{ik}$. The tuning parameter $\Delta>0$ is chosen on a prespecified grid to maximize
\begin{equation}
J(\Delta)
=
\frac{1}{n_1n_0}
\sum_{i:Y_i=1}\sum_{j:Y_j=0}
\ind\!\left\{S_i(\Delta) > S_j(\Delta)\right\}
\label{eq:rounding}
\end{equation}
 subject to the constraint 
$$ 
 w_k(\Delta)\in \{1, \cdots, L\}  \Leftrightarrow \max\left\{\frac{1}{L+0.5}\max_{\beta_k^{\mathrm{logit}}>0} \beta_k^{\mathrm{logit}}, -2\min_{\beta_k^{\mathrm{logit}}<0} \beta_k^{\mathrm{logit}}\right\}\le \Delta \le 2\max_{\beta_k^{\mathrm{logit}}>0} \beta_k^{\mathrm{logit}}.$$
 We also can account for this constraint by adding a penalty term of the form, $\lambda\left\{\ind(\max_k\mathrm{round}(\beta_k^{\mathrm{logit}}/\Delta)>L)+\ind(\min_k\mathrm{round}( \beta_k^{\mathrm{logit}}/\Delta)<0)\right\}$.
 This benchmark is important because it reflects how integer scoring rules are often built in practice. 

\section{Simulation Studies}
\label{sec:simulation}

\subsection{Simulation design}

We evaluated the methods in Section~\ref{sec:method} under three simulation settings. The primary estimator was the base algorithm \ref{alg:greedy}, which we compared with its local-step and look-ahead variants, the local-step look-ahead variant, and the logistic-regression-based integer scoring benchmark. In all settings, the greedy procedures were initialized at $\bbeta^{(0)}=\bzero$ and restricted to the binary action set $\cA=\{0,1\}$. For the benchmark method, we set $\lambda=1$ and fixed the largest admissible integer coefficient at $c_0=1$. Performance was evaluated by the test-set AUC on an independent sample of size $n_{\mathrm{test}}=5000$. For each setting, we considered training sample sizes $n\in\{100,200,400\}$ and averaged the results over $B=1000$ replications. In a separate study, we compared the computational speed of the different methods under Setting 1 while varying one design parameter at a time. Specifically, we considered $L\in \{1,2,3,4,5\}$, $p\in \{5,10,15,20,25\}$, and $n\in \{100,200,300,400,500\}$, using $L=1$, $p=20$, and $n=200$ as the baseline values for the parameters not being varied.


\paragraph{Setting 1: Independent sparse-signal logistic design.}
Setting~1 serves as a simple baseline with a sparse monotone signal. We generate latent variables $\tilde{X}_1,\ldots,\tilde{X}_{20}$ independently from the standard normal distribution and define the observed binary predictors by $X_j=\ind( \tilde{X}_j>0)$. Conditional on $\bX=(X_1,\dots,X_{20})$, the outcome is generated according to
$\Pr(Y=1\mid\bX)=\mbox{expit}\!\left(-2 + \sum_{j=1}^{6} X_j\right).$
Under this model, the six signal predictors contribute equally to the outcome, while the remaining 14 predictors have no effect. This setting provides a transparent benchmark for assessing whether each method can recover a sparse score when the data-generating mechanism is well aligned with the discrete logistic model.

\paragraph{Setting 2: Correlated Gaussian-to-binary design with outliers.}
Setting~2 creates a correlated contaminated Gaussian-to-binary design. It begins by drawing $19$ latent predictors from a mean-zero multivariate normal distribution with unit variance and exchangeable correlation $\rho=0.9$. We then generate $\tilde{X}_{20} = \left(\sum_{j=1}^{19} \tilde{X}_{j} \right)\times\varepsilon$ with $\varepsilon \sim N(1,1)$, and let $ Y = \ind\left(\sum_{j=1}^{19} \tilde{X}_{j}\ge 0\right).$ 
Furthermore, to introduce contamination, for all observations, with a 25\% chance, we replace $\tilde{X}_{1}, \cdots, \tilde{X}_{20}$ and $Y$ by independent fresh draws from $N(-1,1)$ and 1, respectively. 
Finally, the observed binary predictor $X_{j}=\ind(\tilde{X}_{j}>0).$ 
These outliers tend to have small $\eta$ and therefore resemble low-risk subjects, but have an outcome $Y=1.$ Consequently, the setting tests whether a method can still identify useful structure under strong dependence and in the presence of a subgroup of outliers. 

\paragraph{Setting 3: Signal--decoy design with outliers.}
Setting 3 is designed so that the strongest apparent signal comes from the signal block, while a smaller positive subgroup is identified mainly by the joint decoy pattern and individual decoy features are not informative on their own. It tests whether a method can move beyond the easiest signal path and recover a weaker but still useful alternative structure. The detailed simulation setting is given in Appendix B.

\subsection{Simulation results}

Figure~\ref{fig:auc_boxplots} and Table~\ref{tab:sim_results} in Appendix C summarize the test-set AUC results across 1000 replicates. In Setting~1, all methods improve with sample size, and the four discrete search procedures are nearly indistinguishable. The greedy and local-step variants are numerically identical up to rounding, and the look-ahead variants differ only trivially from them. This is consistent with a data-generating mechanism that is already well aligned with a simple additive binary score. Logistic + tuned rounding is clearly weaker at $n=100$ (mean AUC $0.67$ versus about $0.71$ for the discrete methods), but the gap narrows as $n$ increases and essentially disappears by $n=400$.

In Setting~2, all four discrete optimization procedures clearly outperform both logistic baselines at every sample size. Their mean AUCs range from about $0.86$ to $0.88$, compared with $0.69$, $0.75$, and $0.82$ for logistic + tuned rounding and $0.78$, $0.84$, and $0.86$ for logistic regression at $n=100$, $200$, and $400$. Figure~\ref{fig:auc_boxplots} also shows substantially tighter dispersion for the discrete methods, especially relative to the tuned-rounding benchmark, indicating greater stability under strong correlation and contamination. The four discrete procedures remain close, although the look-ahead variants retain a small but consistent advantage.

Setting~3 is the most challenging and the one in which look-ahead helps most. The standard greedy and local-step methods remain near $0.64$--$0.65$, whereas the look-ahead variants improve from $0.66$ at $n=100$ to $0.68$ at $n=400$. Logistic + tuned rounding performs worst throughout, with mean AUCs from $0.52$ to $0.59$, and even continuous logistic regression remains below the look-ahead procedures at every sample size. This pattern matches the design: a myopic rule is drawn toward the easier marginal signal, while look-ahead is better able to recover the weaker multi-step rescue structure.

\begin{figure}[ht]
\centering
\includegraphics[width=0.65\linewidth]{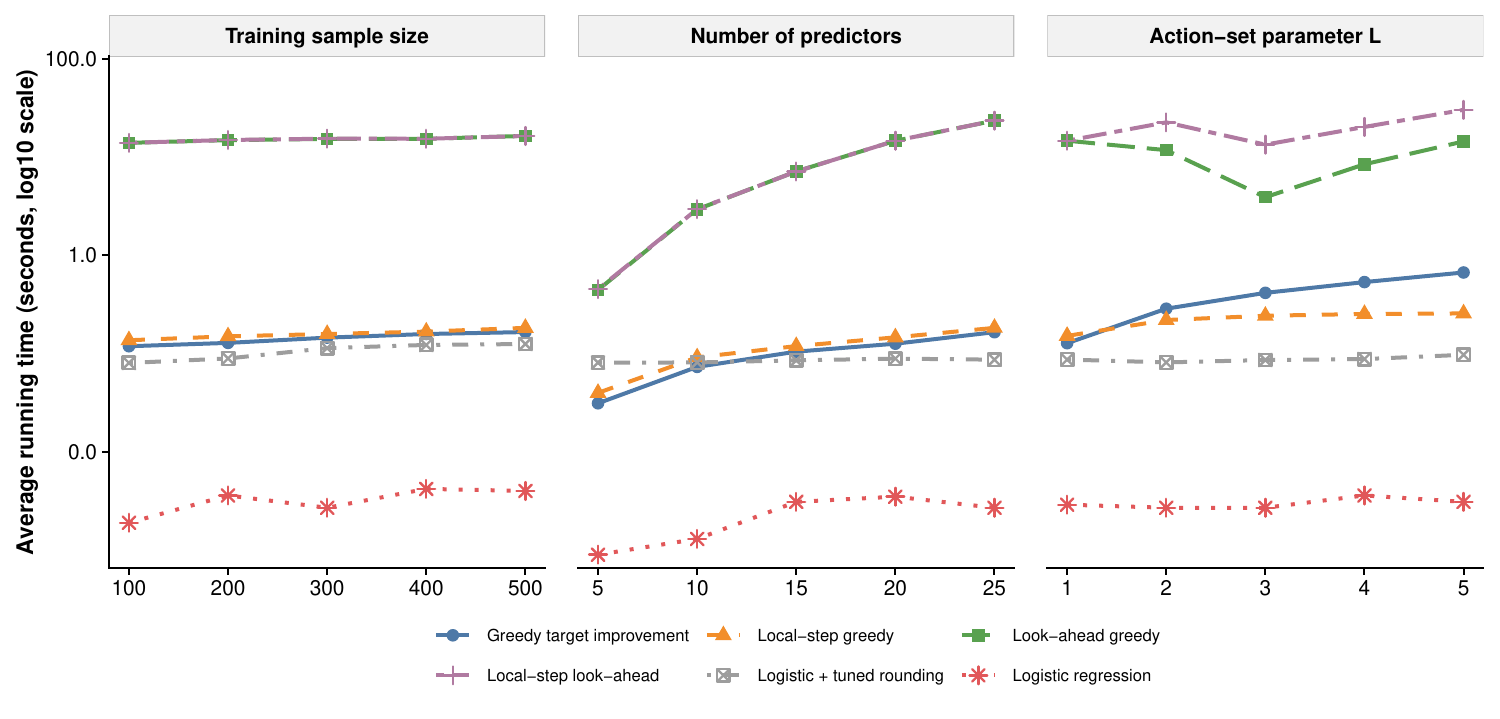}
\caption{Average running time across 50 replicates for different cases.}
\label{fig:running_time}
\end{figure}

Lastly, Figure~\ref{fig:running_time} summarizes average running times on a log scale. From the figure, we can see that look-ahead methods are always the most computationally expensive, whereas logistic regression is the fastest. Running time changes slightly with training sample size over the range considered, increases substantially with the number of predictors, and shows a more modest dependence on $L$. This can be explained by the phenomena that larger $p$ and $L$ increase not only the cost for searching each update but also the number of updates required to reach the final solution.

\begin{figure}[ht]
\centering
\includegraphics[width=0.9\linewidth]{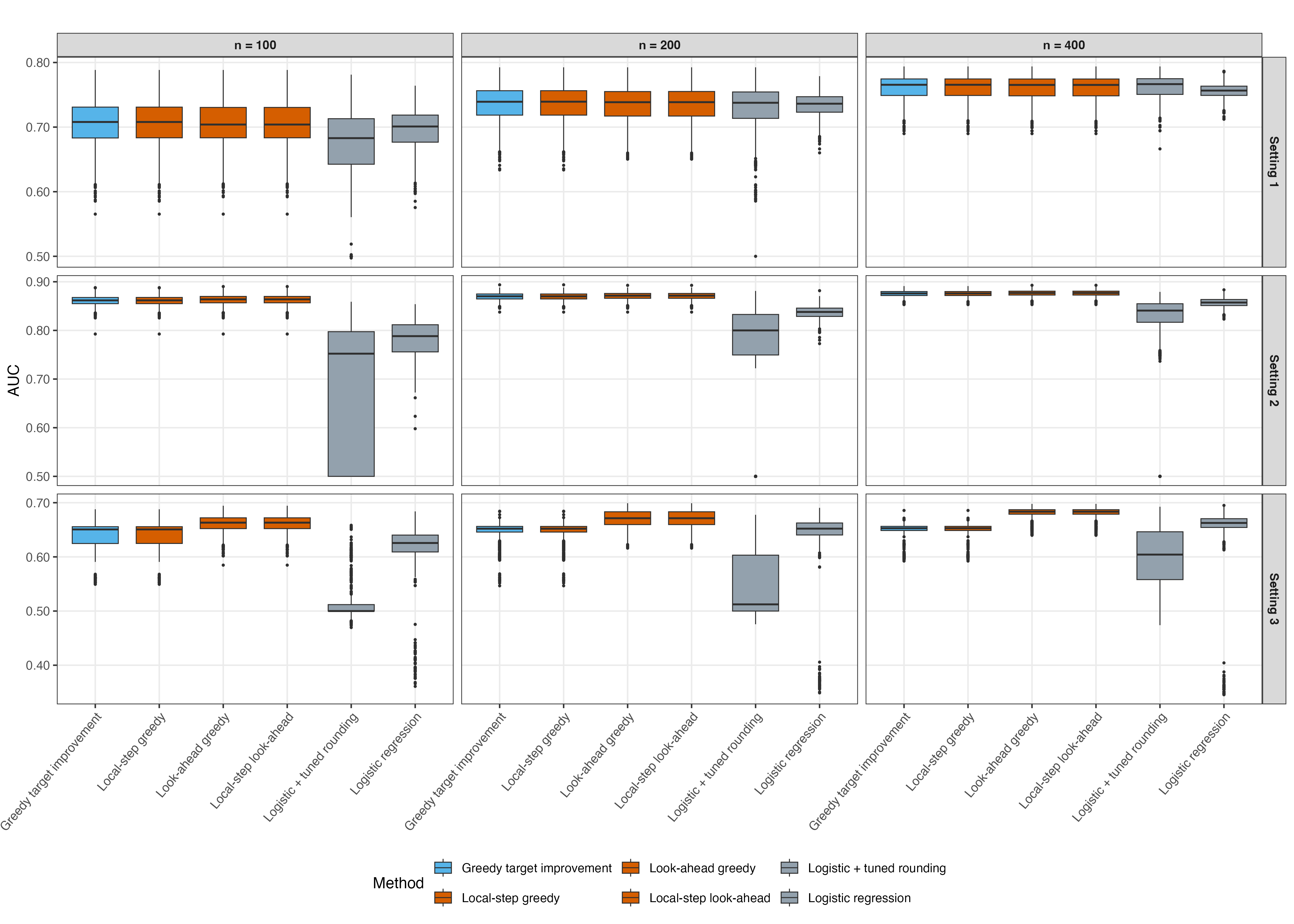}
\caption{Average test-set AUC across 1000 replicates for different methods.}
\label{fig:auc_boxplots}
\end{figure}

\section{Real Application}

We applied the proposed methods to the 1\% SneakPeek sample from Epic Cosmos, with analysis focused on $40{,}622$ patients older than 65 years. The primary outcome was 180-day mortality. Predictors consisted of 22 disease conditions derived from the CCSR ICD-10-CM diagnosis categories, together with three age-group indicators and one sex indicator. Age was represented by the three indicators $\ind\{\mathrm{Age}\in[65,75)\}$, $\ind\{\mathrm{Age}\in[75,85)\}$, and $\ind\{\mathrm{Age}\in[85,100]\}$, and sex was represented by $\ind\{\mathrm{Sex}=\mathrm{Male}\}$. We split the cohort into a training set of $24{,}373$ patients ($60\%$) and a test set of $16{,}249$ patients ($40\%$), fit each method on the training set, and evaluated discrimination on test set.


{Table~\ref{tab:real_results} summarizes test-set AUCs in the real application. As in the simulation study, the local-step variants coincided with their corresponding base procedures. Specifically, the greedy and local-step greedy methods both achieved an AUC of $0.69$, whereas the look-ahead and local-step look-ahead methods both achieved an AUC of $0.70$. The Charlson Comorbidity Index \citep{charlson1987} attained an AUC of $0.66$, and logistic regression with tuned rounding performed poorly, with an AUC of $0.50$, indicating that post hoc discretization can fail badly. Overall, these results suggest that direct optimization yields clinically interpretable discrete scores with competitive discrimination ability, while look-ahead provides a modest improvement over the greedy base procedure.}

\begin{figure}[ht]
\centering
\includegraphics[width=0.45\linewidth]{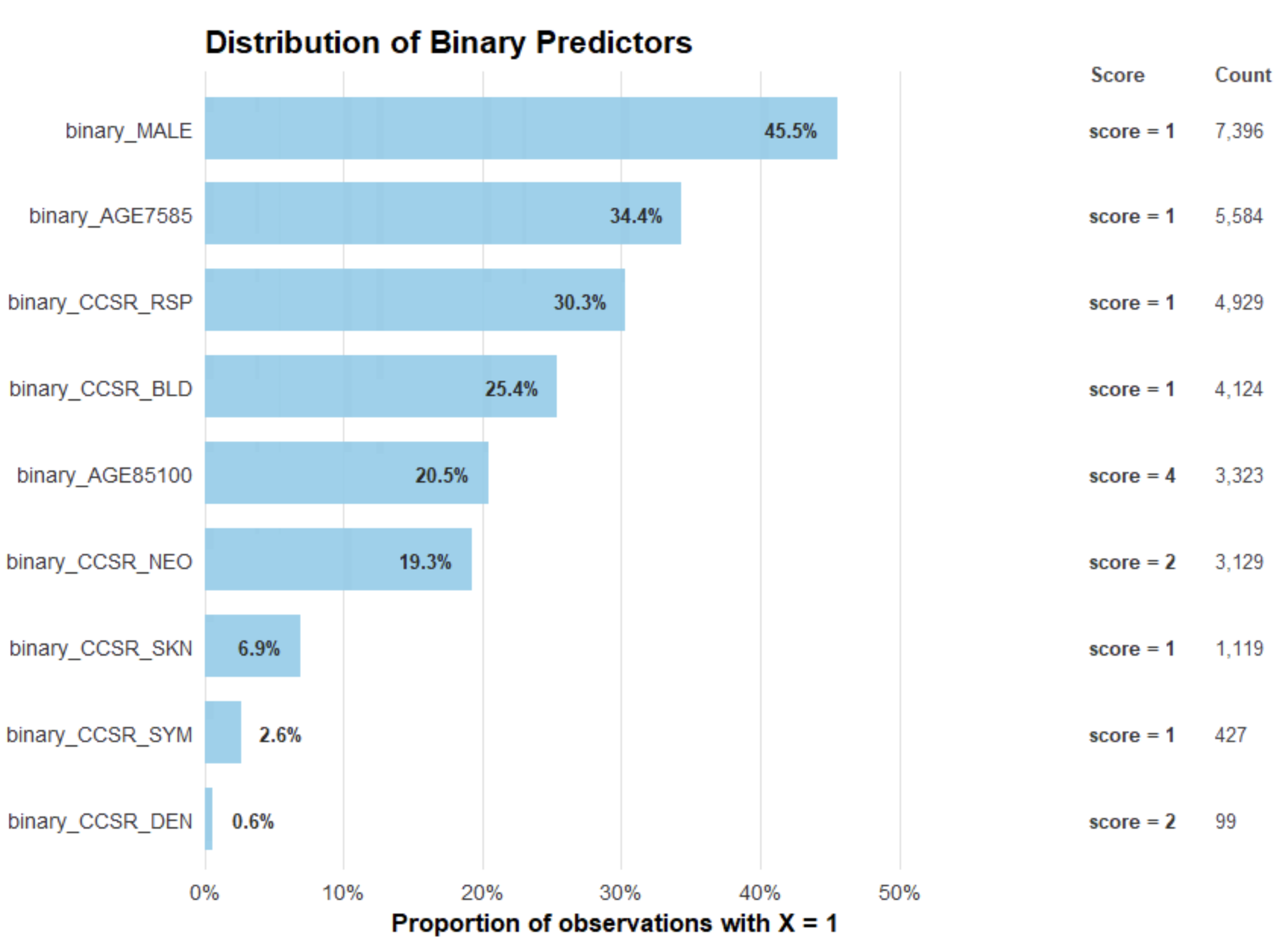}\hfill
\includegraphics[width=0.45\linewidth]{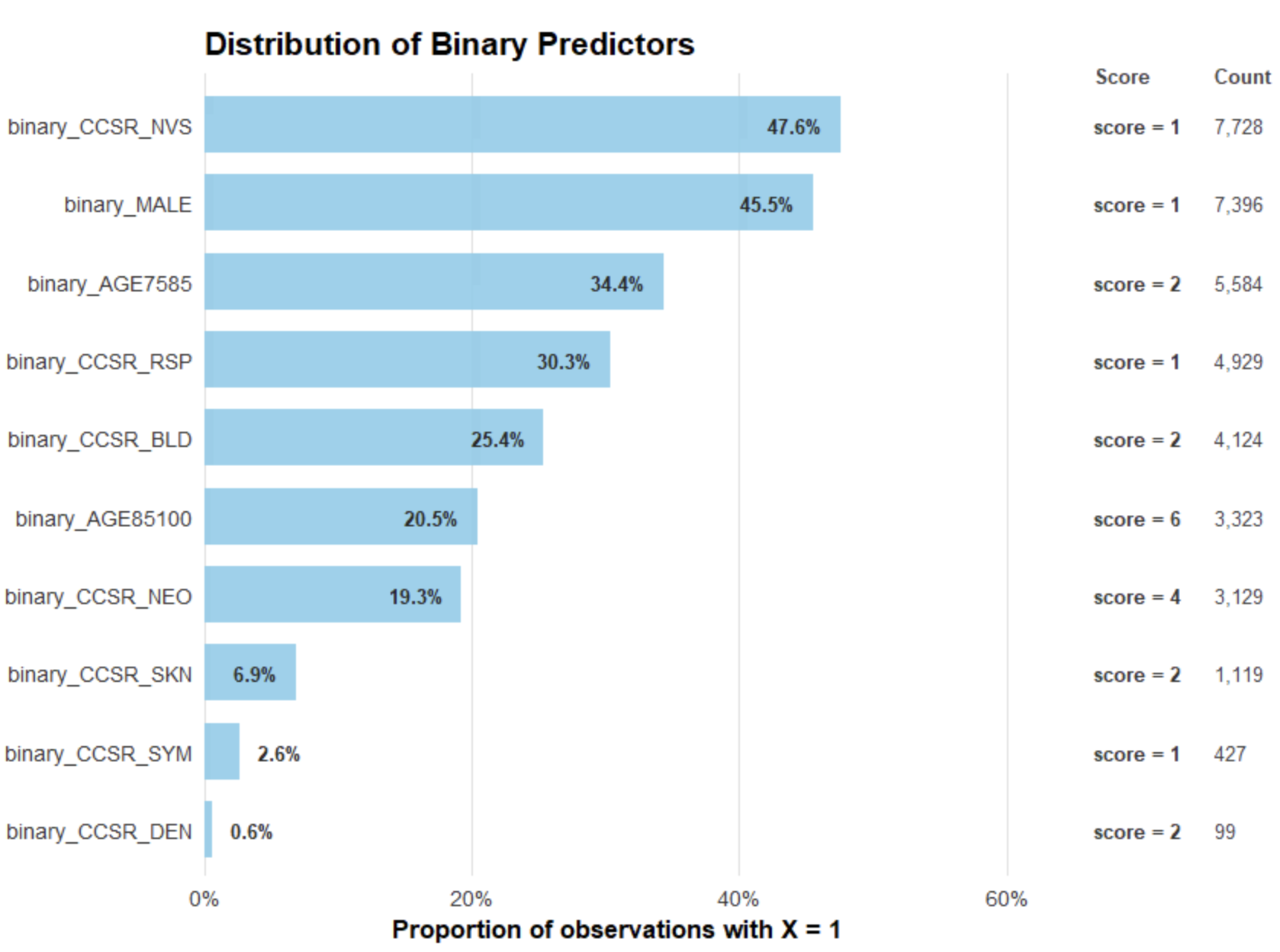}
\caption{Learned scores for basic greedy score (left) and look-ahead score (right).}
\label{fig:real_score_weights}
\end{figure}

\begin{figure}[ht]
\centering
\includegraphics[width=0.45\linewidth]{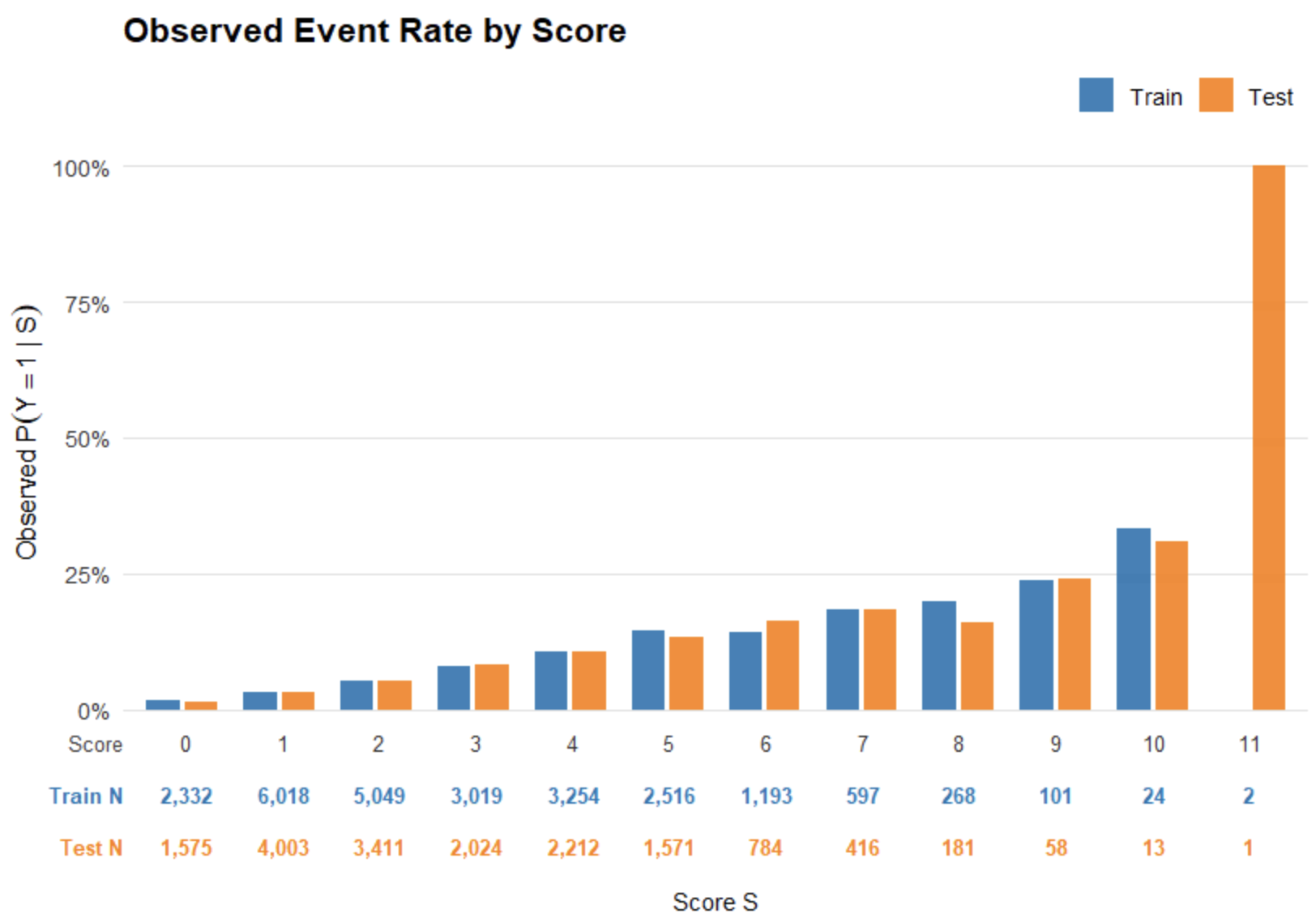}\hfill
\includegraphics[width=0.45\linewidth]{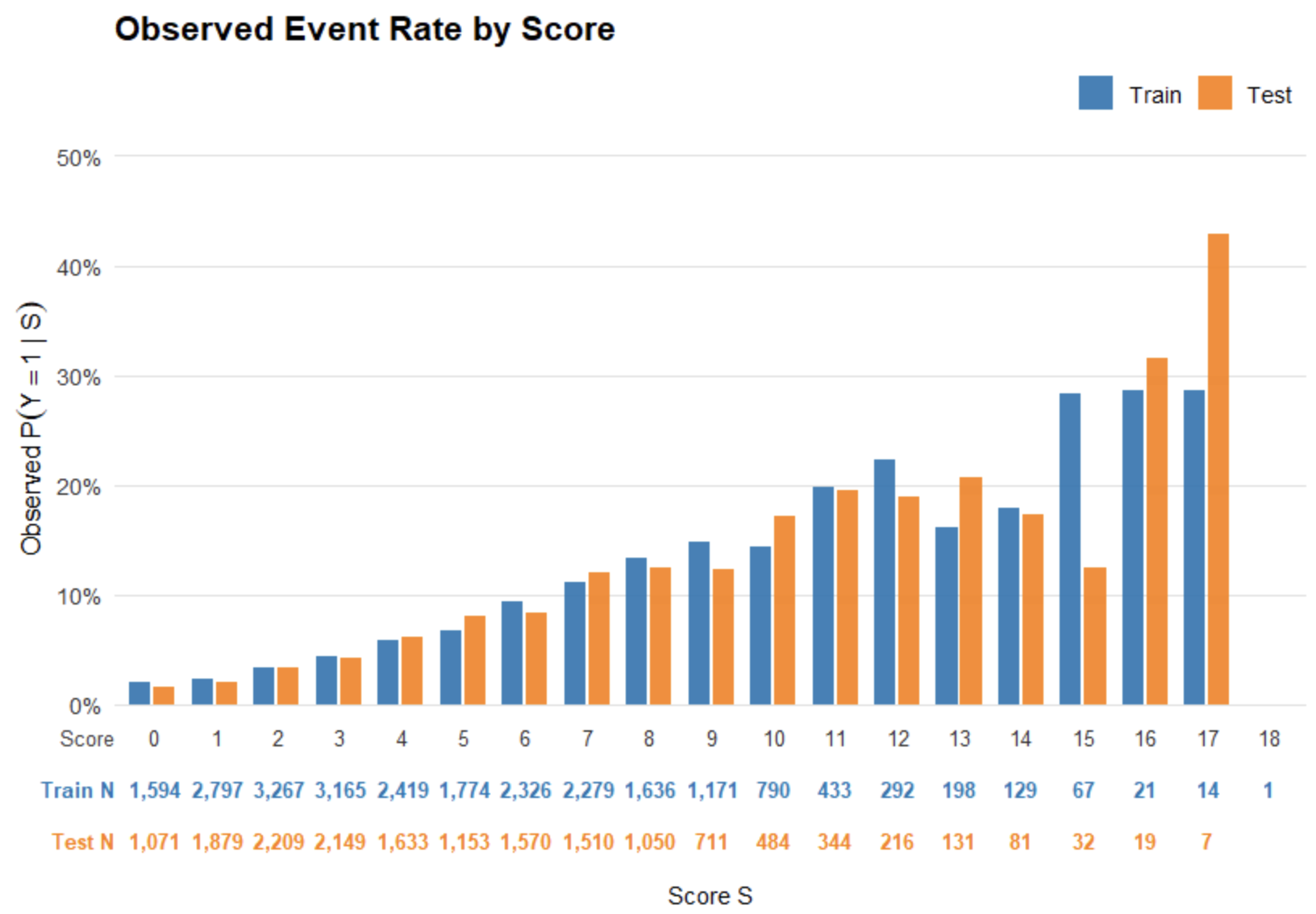}\\[4pt]
\caption{Observed event rate by score for basic greedy score (left) and look-ahead score (right).}
\label{fig:real_score_performance}
\end{figure}

Figures~\ref{fig:real_score_weights} and \ref{fig:real_score_performance} summarize the learned scores and their empirical performance. The basic greedy score selected 9 active predictors, with the largest weight on age 85--100 and additional contributions from sex, age 75--85, respiratory disease, blood disease, neoplasms, skin disease, symptoms, and dental disease. The look-ahead score selected 10 predictors and produced a richer weighting scheme, adding a nervous-system indicator and assigning larger weights to age 85--100, neoplasms, and blood disease. In both cases, the observed event rate increased with the score in the test set, indicating meaningful monotone risk stratification. The increase is more gradual and extends over a wider range for the look-ahead score, which is consistent with its slightly better discrimination. The occational violation of the monotone trend occurs between small strata likely due to random fluctuation. 


\begin{table}[H]
\caption{Summary test-set AUC results for the Epic Cosmos SneakPeek analysis.}
\label{tab:real_results}
\centering
\footnotesize
\begin{tabular}{lcccccc}
\toprule
 & \shortstack{Greedy target\\improvement} & \shortstack{Local-step\\greedy} & \shortstack{Look-ahead\\greedy} & \shortstack{Local-step\\look-ahead} & \shortstack{Logistic +\\rounding} & \shortstack{Charlson\\index} \\
\midrule
Test AUC & 0.69 & 0.69 & 0.70 & 0.70 & 0.50 & 0.66 \\
\bottomrule
\end{tabular}
\end{table}

\section{Discussion}

This paper studies how to learn interpretable point-based clinical risk scores, which is equivalent to developing regression models with discrete-valued coefficients. 
{Our method is interpretable in the practical sense, as it can provide transparent, sparse, monotone, and manually computable point-based rules.}
The central idea is to optimize the score directly in the same constrained coefficient space, rather than fit a continuous model and discretize it afterward. The objective function can be as complex as the c-index for a binary outcome, which is neither continuous nor convex. The proposed greedy search algorithm is easy to implement, and can be further extended using reinforcement learning. 
{Finally, our approach provides a general framework in which different objective functions can be considered. For example, the same framework could be adapted to longitudinal or time-to-event settings using concordance criteria such as Harrell's c-index, Uno's c-index, or time-dependent c-index formulations \citep{harrell1996,uno2011,heagerty2005}.}
Several limitations should be noted and warrant further research. First, the greedy procedures remain search algorithms and therefore may converge to local optima. Second, the computational burden of further reinforcement-learning extensions can be substantial.

\section*{Acknowledgement}
This work is supported by the US National Institutes of Health grants R01HL089778 (LT).
Data used in this study came from Epic Cosmos, a dataset created in collaboration with a community of health systems using Epic representing more than 301 million patient records from over 2,068 hospitals and 47.1K clinics as of May, 2026. The community represents patients from all 50 states, D.C., Canada, Lebanon, and Saudi Arabia.

\newpage

\bibliographystyle{plainnat}
\bibliography{paper_refs}

\setcounter{equation}{0} 
\renewcommand{\theequation}{A\arabic{equation}}

\newpage
\setcounter{page}{1} 
\section*{Appendix A: Effective Search Complexity}

Let \({\mathcal A}_p=\{0,1,\ldots,L\}^p\), so that
\(M=|{\mathcal A}_p|=(L+1)^p\). Because \(\mathcal A\) is finite, there is no
continuous local parameter dimension in the usual parametric sense. Instead,
the relevant notion of complexity comes from searching over the
\(M=(L+1)^p\) candidate scoring rules, whose logarithmic search complexity is
\(\log M=p\log(L+1)\).

Under the null hypothesis \(Y\perp \bX\), every fixed
\(\bbeta\in\mathcal A\) has population AUC equal to \(1/2\). However,
maximizing the empirical AUC over \(\mathcal A\) still introduces
selection-induced optimism analogous to overfitting. This motivates the definition of null AUC optimism as summarized in Definition \ref{def:1}.

\begin{definition}[Null AUC optimism]
\label{def:1}
With \({\mathcal A}_p\) as defined above, let \(\ell(\bbeta)\)
given in (\ref{eq:target}) denote the empirical AUC of the score \(S_{\bbeta}(\bX)\) for each
\(\bbeta\in\mathcal A\), and let
\[
\hat\bbeta
=
\arg\max_{\bbeta\in\mathcal A}
\ell(\bbeta).
\]
We define the null AUC optimism of the selection procedure by
\[
\mathrm{Opt}_0
=
E_0\left[\ell(\hat\bbeta)\right]-\frac{1}{2},
\]
where \(E_0\) denotes expectation under the null hypothesis \(Y\perp \bX\).
\end{definition}

The corresponding optimism-based measure of AUC search complexity is defined in Definition \ref{def:2}.

\begin{definition}[AUC effective search complexity]
\label{def:2}
Let \(\mathrm{Opt}_0\) denote the null AUC optimism of the selection
procedure. Let
\[
\sigma_{\mathrm{AUC},0}^2
=
\frac{1}{|\mathcal A|}
\sum_{\bbeta\in\mathcal A}
\operatorname{Var}_0\left\{
\ell(\bbeta)
\right\}
\]
denote the average null variance of the empirical AUC over all candidate rules. We define the AUC effective search complexity by
\[
\mathrm{ESC}_{\mathrm{AUC}}
=
\frac{\mathrm{Opt}_0^2}{2\sigma_{\mathrm{AUC},0}^2}.
\]
\end{definition}

This quantity is introduced as an optimism-based measure of the effective complexity induced by searching over the finite class \({\mathcal A}_p\). Larger values of \(\mathrm{ESC}_{\mathrm{AUC}}\) indicate greater procedural flexibility, meaning that the method has more capacity to adapt to random fluctuations in the training labels and thereby overfit the empirical AUC.

Estimating \(\mathrm{Opt}_0\) is nontrivial because it depends on the distribution of the maximum empirical AUC over \(\mathcal A\), and hence on the dependence structure among the candidate scoring rules.
We next outline heuristic derivations for the optimism in two canonical settings, with \(\bX\) following a Gaussian distribution in one case and a Bernoulli distribution in the other. Although the main paper focuses on binary predictors, we begin with the Gaussian case because it yields a cleaner and more intuitive approximation, which helps clarify the logic before turning to the binary setting.

\subsection*{Approximate AUC Effective Search Complexity with Canonical Setting 1}

First assume that \(Y_i\sim \mathrm{Bernoulli}(1/2)\),
\(\bX_i=(X_{i1},\ldots,X_{ip})^\top\sim N(0,I_p)\), and \(Y_i\perp \bX_i\) for $i=1,\dots,n$. Let
\(n_1=\sum_{i=1}^n I(Y_i=1)\) and \(n_0=\sum_{i=1}^n I(Y_i=0)\) denote the numbers of
cases and controls, and assume \(n_1=n_0=n/2\). Let \(\bbeta \in {\mathcal A}_p\) and define the linear score
\(S_{\bbeta}(X)=\bX^\top\bbeta\) with the empirical AUC $\ell(\bbeta)$ in (\ref{eq:target}).
The AUC-based effective search complexity admits the following heuristic asymptotic approximation:
\[
\mathrm{ESC}_{\mathrm{AUC}}
\approx
p\left[
1-
\frac{6}{\pi}
\arcsin\left\{\frac{3L}{4(2L+1)}\right\}
\right]\log(L+1).
\]

We now outline the derivation underlying the preceding approximation.
For any fixed nonzero \(\bbeta\), since \(\bX^\top\bbeta\) is continuous normal, 
\[
\sigma_{\mathrm{AUC},0,\mathrm{cont}}^2=\operatorname{Var}_0\{
\ell(\bbeta)
\}
\approx
\frac{1}{3n}.
\]
Under the Gaussian null model,
\[
S_{\bbeta}(\bX)=\bX^\top\bbeta,
\qquad 
S_{\bgamma}(\bX)=\bX^\top\bgamma
\]
are jointly normal with correlation \(\rho_{\bbeta\bgamma}=\frac{\bbeta^\top\bgamma}{\|\bbeta\|\,\|\bgamma\|}\).
The leading covariance between the two empirical AUCs is
\[
\operatorname{Cov}_0
\left\{
\ell(\bbeta),
\ell(\bgamma)
\right\}
\approx
\frac{2}{n\pi}
\arcsin\left(\frac{\rho_{\bbeta\bgamma}}{2}\right).
\]
 The average off-diagonal covariance is
\[
\bar \Sigma_{\mathrm{cont}}
\approx
\frac{2}{nM(M-1)\pi}
\sum_{\bbeta\ne\bgamma}
\left[
\arcsin\left(
\frac{\rho_{\bbeta\bgamma}}{2}
\right)
\right],
\]
and the average correlation coefficient among candidate empirical AUCs is approximately
\begin{equation}
\bar R_{\mathrm{cont}}
\approx
\frac{6}{\pi M(M-1)}
\sum_{\bbeta\ne\bgamma}
\left[
\arcsin\left(
\frac{\rho_{\bbeta\bgamma}}{2}
\right)
\right]. \label{eq:meancorr}
\end{equation}

A further simplification is to replace the random correlation
\(
\rho_{\bbeta\bgamma}
\)
in (\ref{eq:meancorr}) by its ``limit'' as $p \rightarrow \infty.$  Specifically, for large \(p\), by the law of large numbers,
\[
\frac{\bbeta^\top\bgamma}{p}
\approx
\{E(B)\}^2,
\quad
\frac{\|\bbeta\|^2}{p}
\approx
E(B^2),
\qquad
\frac{\|\bgamma\|^2}{p}
\approx
E(B^2),
\]
and 
$$\rho_{\bbeta\bgamma} \approx 
\frac{3L}{2(2L+1)}.$$
In summary, a simple approximation to the average AUC correlation is
\begin{equation}
\bar R_{\mathrm{cont}}
\approx
\frac{6}{\pi}
\arcsin\left[
\frac{3L}{4(2L+1)}
\right].
\label{eq:A1}
\end{equation}
From \eqref{eq:A1}, it is easy to see that the average AUC correlation depends on \(L\). As such, we write this dependence explicitly as
\[
\bar R_{\mathrm{cont}}=\bar R_{\mathrm{cont}}(L).
\]
Using an average-correlation approximation, we summarize the dependence among the \(M\) empirical AUC processes by an average pairwise correlation \(\bar R_{\mathrm{cont}}(L)\). We then define an effective number of candidate scoring rules by
\[
M_{\mathrm{eff}} \approx M^{1-\bar R_{\mathrm{cont}}(L)}.
\]
This quantity could be interpreted as the number of approximately independent candidate scoring rules that would produce a similar extreme-value behavior to the original correlated collection.
More explicitly, the approximation replaces the maximum of a correlated Gaussian vector
\[
(Z_1,\ldots,Z_M)^\top \sim N(0,\Sigma_M)
\]
by the maximum of an independent Gaussian vector
\[
(\widetilde Z_1,\ldots,\widetilde Z_{M_{\mathrm{eff}}})^\top \sim N(0,I_{M_{\mathrm{eff}}}),
\]
in the sense that
\[
\max_{1\le m\le M} Z_m
\quad\text{can be approximated by}\quad
\max_{1\le m\le M_{\mathrm{eff}}} \widetilde Z_m.
\]
To motivate the above definition, if the dependence structure is approximated by an equicorrelation matrix
\[
\Sigma_{M,\rho} = (1-\rho)I_M+\rho \mathbf 1_M\mathbf 1_M^\top,
\]
with off-diagonal correlation \(\rho\), then larger \(\rho\) reduces the effective number of independent candidate rules. This approximation of \(M_{\mathrm{eff}}\) uses the average correlation \(\bar R_{\mathrm{cont}}(L)\) as a scalar summary of \(\rho\).
These lead to
\begin{equation}
M_{\mathrm{eff}}
\approx (L+1)^{p[1-\bar{R}_{\mathrm{cont}}(L)]}
\approx
(L+1)^{
p\left\{1-\frac{6}{\pi}
\arcsin\left[
\frac{3L}{4(2L+1)}
\right]\right\}}
\label{eq:A2}
\end{equation}
The correlation-adjusted null optimism and the corresponding AUC effective search complexity can then be approximated by plugging in \eqref{eq:A2} as
\[
\mathrm{Opt}_0
\approx
\sqrt{
\frac{2\log M_{\mathrm{eff}}}{3n}
}.
\]
and
\begin{equation}
\mathrm{ESC}_{\mathrm{AUC}}
\approx
\log M_{\mathrm{eff}}
\approx p\left[1 - \frac{6}{\pi} \arcsin\left\{\frac{3L}{4(2L+1)}\right\}\right]\log(L+1),
\label{eq:A3}
\end{equation}
respectively.

\subsection*{Approximate AUC Effective Search Complexity with Canonical Setting 2}

The current paper focuses on the setting that \(X_{i1},\ldots,X_{ip}, Y_i\stackrel{\mathrm{i.i.d.}}{\sim} \mathrm{Bernoulli}(1/2)\). A heuristic asymptotic approximation to the AUC effective search complexity is
\[
\mathrm{ESC}_{\mathrm{AUC}}^{\mathrm{bin}}
\approx
p\{1-\bar R_{\mathrm{bin}}(p,L)\}\log(L+1).
\]
Under the additional large-\(p\), negligible-tie approximation, this further simplifies to
\[
\mathrm{ESC}_{\mathrm{AUC}}^{\mathrm{bin}}
\approx
p\left[
1-
\frac{6}{\pi}
\arcsin\left\{\frac{3L}{4(2L+1)}\right\}
\right]\log(L+1).
\]

To derive this approximation, we follow the same general strategy as in canonical setting 1, but replace \eqref{eq:A1} with its binary analogue.
For the covariance calculation, define the tie-adjusted AUC kernel
\[
h_{\bbeta}(\bx,\bz)
=
\ind\{S_{\bbeta}(\bx)>S_{\bbeta}(\bz)\}
+
\frac12 \ind\{S_{\bbeta}(\bx)=S_{\bbeta}(\bz)\}.
\]
Then for \(\bZ\) as an independent copy of \(\bX\), we have
\[
q_{\bbeta}(\bx)=E_{\bZ}\{h_{\bbeta}(\bx,\bZ)\}.
\]
Under the null hypothesis, \(E_{\bX}\{q_{\bbeta}(\bX)\}=1/2\). For two fixed
candidate rules \(\bbeta\) and \(\bgamma\), the leading covariance between their empirical AUCs is 
\[
\operatorname{Cov}_0
\left\{
\ell(\bbeta),
\ell(\bgamma)
\right\}
\approx
\frac4n
\left[
E_{\bX}\{q_{\bbeta}(\bX)q_{\bgamma}(\bX)\}
-\frac14
\right].
\]

Define
\[
A_{\mathrm{bin}}
=
E_{\bX}\{m(\bX)^2\}-\frac14,
\qquad
D_{\mathrm{bin}}
=
E_{\bbeta}E_{\bX}\{q_{\bbeta}(\bX)^2\}-\frac14.
\]
We have
\[
\frac{1}{M(M-1)}
\sum_{\bbeta\ne\bgamma}
\left[
E_{\bX}\{q_{\bbeta}(\bX)q_{\bgamma}(\bX)\}
-\frac14
\right]
=
\frac{M A_{\mathrm{bin}}-D_{\mathrm{bin}}}{M-1}.
\]
Hence the average off-diagonal covariance is
\[
\bar\Sigma_{\mathrm{bin}}
\approx
\frac4n
\left[
\frac{M A_{\mathrm{bin}}-D_{\mathrm{bin}}}{M-1}
\right],
\]
 the average fixed-rule null variance is
\[
\sigma_{\mathrm{AUC},0,\mathrm{bin}}^2
\approx
\frac4n D_{\mathrm{bin}},
\]
and the average correlation coefficient is
\[
\bar R_{\mathrm{bin}}
\approx
\frac{
M A_{\mathrm{bin}}-D_{\mathrm{bin}}
}{
(M-1)D_{\mathrm{bin}}
}.
\]
For large \(M\), this simplifies to
\[
\bar R_{\mathrm{bin}}
\approx
\frac{
E_{\bX}\{m(\bX)^2\}-\frac14
}{
E_{\bbeta}E_{\bX}\{q_{\bbeta}(\bX)^2\}-\frac14
},
\]
where the expectation over \(\bbeta\) is taken with respect to the uniform distribution on
\(
\{0,1,\ldots,L\}^p.
\)
 For a fixed coefficient vector
\(\bbeta\), let
\[
p_{\bbeta,s}=P_{\bX}\{S_{\bbeta}(\bX)=s\},
\]
and 
\[
q_{\bbeta,s}
=
\sum_{u<s}p_{\bbeta,u}+\frac12p_{\bbeta,s}.
\]
Then
\[
E_{\bX}\{q_{\bbeta}(\bX)^2\}
=
\sum_s p_{\bbeta,s}q_{\bbeta,s}^2
=
\frac13-\frac1{12}\sum_s p_{\bbeta,s}^3,
\]
and
\[
D_{\mathrm{bin}}
=
E_{\bbeta}E_{\bX}\{q_{\bbeta}(\bX)^2\}-\frac14
=
\frac1{12}
\left[
1-
E_{\bbeta}\left\{\sum_s p_{\bbeta,s}^3\right\}
\right].
\]
Therefore the leading null variance of a fixed randomly chosen AUC rule is
\[
\sigma_{\mathrm{AUC},0,\mathrm{bin}}^2
\approx
\frac4nD_{\mathrm{bin}}
=
\frac1{3n}
\left[
1-
E_{\bbeta}\left\{\sum_s p_{\bbeta,s}^3\right\}
\right]
\]
which is $1/(3n)$, when \(p\) is moderately large.  To approximate the average off-diagonal covariance, we let
\(\bbeta\) and \(\bgamma\) be two independently drawn coefficient vectors from
\(\{0,1,\ldots,L\}^p\). Conditional on \(\bX=\bx\), the quantities
\(q_{\bbeta}(\bx)\) and \(q_{\bgamma}(\bx)\) are independent over
\(\bbeta,\bgamma\), and hence the off-diagonal term is governed by
\[
E_{\bX}\{m(\bX)^2\}-\frac14,
 ~\mbox{where}~
m(\bx)=E_{\bbeta}\{q_{\bbeta}(\bx)\}.
\]
By exchangeability, \(m(\bx)\) depends on \(\bx\) only through
\[
R=\sum_{j=1}^p x_j.
\]
and we may
write
\(
m(\bx)=R_L(r),
\)
where
\[
R_L(r)
=
P\left(
\sum_{\ell=1}^r B_\ell W_\ell
>
\sum_{\ell=1}^{p-r}B'_\ell W'_\ell
\right)
+
\frac12
P\left(
\sum_{\ell=1}^r B_\ell W_\ell
=
\sum_{\ell=1}^{p-r}B'_\ell W'_\ell
\right),
\]
with \(r=\sum_{j=1}^p x_j\),
\(B_\ell,B'_\ell\stackrel{iid}{\sim}\mathrm{Bernoulli}(1/2)\),
and \(W_\ell,W'_\ell\stackrel{iid}{\sim}\mathrm{Unif}\{0,1,\ldots,L\}\).
Thus, we have
\[
m(\bX)=R_L(R),
 \mbox{ where } 
R\sim\mathrm{Binomial}(p,1/2)
\]
and
\[
A_{\mathrm{bin}}
=
E_{\bX}\{m(\bX)^2\}-\frac14
=
E_R\{R_L(R)^2\}-\frac14.
\]
Consequently, for large \(p\),
\[
\bar R_{\mathrm{bin}}(p,L)
\approx
\frac{A_{\mathrm{bin}}}{D_{\mathrm{bin}}}
=
\frac{
E_R\{R_L(R)^2\}-\frac14
}{
\frac1{12}
\left[
1-
E_{\bbeta}\left\{\sum_s p_{\bbeta,s}^3\right\}
\right]
}
\approx
12\left[E_R\{R_L(R)^2\}-\frac14\right].
\]
We now approximate \(E_R\{R_L(R)^2\}\) for large \(p\). For one coordinate,
\[
BW=
\begin{cases}
0, & B=0,\\
W, & B=1,
\end{cases}
\]
which leads to
\[
E(BW)=\frac L4
\mbox{ and }
\operatorname{Var}(BW)=\frac{L(5L+4)}{48}.
\]
With \(R\sim\mathrm{Binomial}(p,1/2)\), a normal approximation yields
\[
R_L(R)\approx \Phi\left(\frac{3L}{5L+4}Z \right),\mbox{ where }
Z\sim N(0,1).
\]
Therefore, we again obtain
\[
E_R\{R_L(R)^2\}-\frac14
\approx
\frac{1}{2\pi}
\arcsin\left\{
\frac{3L}{4(2L+1)}
\right\}.
\]
and
\[
\bar R_{\mathrm{bin}}(p,L)
\approx
\frac{6}{\pi}
\arcsin\left\{
\frac{3L}{4(2L+1)}
\right\}.
\]
The rest of the derivations follows those in the Gaussian setting.  For small \(p\) and \(L\), \(\mathrm{ESC}_{\mathrm{AUC}}\) can be directly
estimated by Monte Carlo simulation.

\newpage
\section*{Appendix B: Simulation Setting}
Setting~3 creates informative predictors from a signal block (5 predictors) and a decoy block (5 predictors) that is informative only when several decoy features are considered together. 
 Specifically, we first sample $Y\sim \mathrm{Bernoulli}(0.4).$ Then we sample binary predictors as follows.
\begin{enumerate}
\item For $Y=0,$ 5 signal and 10 noise predictors are independently generated by thresholding their corresponding latent variables generated from $N(\bzero_5, \bSigma_{5}(0.9))$ and $N(\bzero_{10}, \bSigma_{10}(0.9)),$ respectively, at zero. 5 decoy binary predictors are generated by a random permutation of $\{1, 0, 0, 0, 0\}.$  
\item For $Y=1,$ 
\begin{itemize}
\item with a probability of 50\%, 5 signal and 10 noise predictors are generated by thresholding $N(\mathbf{2}_5, \bSigma_{5}(0.9))$ and $N(\bzero_{10}, \bSigma_{10}(0.9)),$ respectively. In addition, 5 decoy binary predictors are $\bzero_5;$  
\item with a probability of 40\%, 5 signal and 10 noise predictors are generated by thresholding $N(-\mathbf{2}_5, \bSigma_{5}(0.9))$ and $N(\bzero_{10}, \bSigma_{10}(0.9)),$ respectively. In addition, 5 decoy binary predictors are $\bone_5;$  
\item with a probability of 10\%, 5 signal and 10 noise predictors are generated by thresholding $N(-\bone_5, \bSigma_{5}(0))$ and $N(-\bone_{10}, \bSigma_{10}(0)),$ respectively. In addition, 5 decoy binary predictors are $\bzero_5.$  
\end{itemize}
\end{enumerate}
Here, $\bSigma_{p}(\rho)$ denotes a $p\times p$ matrix with all diagonal and off-diagonal elements being 1 and $\rho,$ respectively, and $\mathbf{j}_p$ denotes a $p-$dimensional vector of $j.$

\newpage

\section*{Appendix C: Simulation Results on AUCs in the Test Set}

\setcounter{table}{0} 
\renewcommand{\thetable}{A\arabic{table}}

\begin{table}[ht]
\caption{Simulation results of different methods under different settings with average test AUC (SD) over $B=1000$ replications on an independent test set of size $5000$.}
\label{tab:sim_results}
\centering
\begin{tabular}{lccc}
\hline
\multicolumn{4}{l}{\textbf{Setting 1: Independent sparse-signal logistic design}} \\
\hline
Method & $n=100$ & $n=200$ & $n=400$ \\
\hline
Greedy target improvement & 0.7057 (0.0366) & 0.7362 (0.0281) & 0.7608 (0.0178) \\
Local-step greedy & 0.7057 (0.0366) & 0.7363 (0.0281) & 0.7608 (0.0178) \\
Look-ahead greedy & 0.7050 (0.0353) & 0.7357 (0.0271) & 0.7604 (0.0181) \\
Local-step look-ahead & 0.7050 (0.0353) & 0.7358 (0.0271) & 0.7604 (0.0181) \\
Logistic + tuned rounding & 0.6697 (0.0604) & 0.7298 (0.0373) & 0.7620 (0.0172) \\
Logistic regression & 0.6961 (0.0310) & 0.7341 (0.0179) & 0.7559 (0.0114) \\
\hline
\multicolumn{4}{l}{\textbf{Setting 2: Correlated Gaussian-to-binary design with outliers}} \\
\hline
Method & $n=100$ & $n=200$ & $n=400$ \\
\hline
Greedy target improvement & 0.8609 (0.0100) & 0.8694 (0.0078) & 0.8757 (0.0061) \\
Local-step greedy & 0.8609 (0.0100) & 0.8693 (0.0078) & 0.8757 (0.0061) \\
Look-ahead greedy & 0.8629 (0.0099) & 0.8706 (0.0076) & 0.8765 (0.0059) \\
Local-step look-ahead & 0.8629 (0.0099) & 0.8706 (0.0076) & 0.8765 (0.0059) \\
Logistic + tuned rounding & 0.6897 (0.1344) & 0.7507 (0.1236) & 0.8157 (0.0807) \\
Logistic regression & 0.7805 (0.0408) & 0.8372 (0.0135) & 0.8572 (0.0095) \\
\hline
\multicolumn{4}{l}{\textbf{Setting 3: Signal--decoy design with outliers}} \\
\hline
Method & $n=100$ & $n=200$ & $n=400$ \\
\hline
Greedy target improvement & 0.6405 (0.0246) & 0.6447 (0.0207) & 0.6483 (0.0162) \\
Local-step greedy & 0.6405 (0.0246) & 0.6447 (0.0207) & 0.6483 (0.0162) \\
Look-ahead greedy & 0.6619 (0.0162) & 0.6704 (0.0150) & 0.6809 (0.0100) \\
Local-step look-ahead & 0.6619 (0.0162) & 0.6704 (0.0150) & 0.6809 (0.0100) \\
Logistic + tuned rounding & 0.5180 (0.0383) & 0.5475 (0.0553) & 0.5938 (0.0592) \\
Logistic regression & 0.6155 (0.0509) & 0.6385 (0.0621) & 0.6512 (0.0588) \\
\hline
\end{tabular}
\end{table}

\newpage
\section*{Appendix D: CCSR Category Definitions}
\label{app:ccsr-definitions}

\setlength{\tabcolsep}{4pt}
\renewcommand{\arraystretch}{1.15}
\begin{longtable}{@{}>{\RaggedRight\arraybackslash}p{0.21\linewidth}>{\RaggedRight\arraybackslash}p{0.06\linewidth}>{\RaggedRight\arraybackslash}p{0.64\linewidth}@{}}
\caption{CCSR diagnosis-category covariates}\label{tab:ccsr-definitions}\\
\toprule
Model covariate & Prefix & Broad category \\
\midrule
\endfirsthead
\caption[]{CCSR diagnosis-category covariates (continued)}\\
\toprule
Model covariate & Prefix & Broad category \\
\midrule
\endhead
\midrule
\multicolumn{3}{r}{\emph{Continued on next page}}\\
\endfoot
\bottomrule
\endlastfoot
\texttt{binary\_CCSR\_CIR} & CIR & Circulatory system \\
\texttt{binary\_CCSR\_END} & END & Endocrine, nutritional, and metabolic disorders \\
\texttt{binary\_CCSR\_DIG} & DIG & Digestive system \\
\texttt{binary\_CCSR\_MUS} & MUS & Musculoskeletal system and connective tissue \\
\texttt{binary\_CCSR\_NVS} & NVS & Nervous system \\
\texttt{binary\_CCSR\_MBD} & MBD & Mental, behavioral, and neurodevelopmental disorders \\
\texttt{binary\_CCSR\_BLD} & BLD & Blood, hematologic, and immune conditions \\
\texttt{binary\_CCSR\_GEN} & GEN & Genitourinary system \\
\texttt{binary\_CCSR\_RSP} & RSP & Respiratory system \\
\texttt{binary\_CCSR\_INJ} & INJ & Injury, poisoning, and other external-consequence conditions \\
\texttt{binary\_CCSR\_EAR} & EAR & Ear and mastoid \\
\texttt{binary\_CCSR\_EYE} & EYE & Eye and adnexa \\
\texttt{binary\_CCSR\_INF} & INF & Infectious and parasitic diseases \\
\texttt{binary\_CCSR\_SKN} & SKN & Skin and subcutaneous tissue \\
\texttt{binary\_CCSR\_NEO} & NEO & Neoplasms \\
\texttt{binary\_CCSR\_MAL} & MAL & Congenital malformations and chromosomal abnormalities \\
\texttt{binary\_CCSR\_SYM} & SYM & Symptoms, signs, and abnormal findings \\
\texttt{binary\_CCSR\_DEN} & DEN & Dental and oral conditions \\
\texttt{binary\_CCSR\_PRG} & PRG & Pregnancy, childbirth, and the puerperium \\
\texttt{binary\_CCSR\_PNL} & PNL & Perinatal conditions \\
\texttt{binary\_CCSR\_FAC} & FAC & Factors influencing health status and health-service contact \\
\texttt{binary\_CCSR\_EXT} & EXT & External causes of morbidity \\
\end{longtable}

\end{document}